\documentclass[conference]{IEEEtran}
\usepackage{amssymb}
\usepackage{cite}
\ifCLASSINFOpdf
  \usepackage[pdftex]{graphicx}
\else
\fi
\usepackage{amsmath}
\usepackage[underline=false]{pgf-umlsd}
\usetikzlibrary{calc}
\usetikzlibrary{arrows.meta}
\usepackage{enumerate}
\usepackage[shortlabels]{enumitem}
\usepackage{algorithmic}

\usepackage{threeparttable}

\usepackage{url}

\usepackage{algorithm}
\usepackage{multicol}
\usepackage{booktabs}
\usepackage{multirow}
\usepackage{makecell}
\usepackage{balance}
\usepackage{siunitx}
\usepackage{tikz}
\newcommand*{\priority}[1]{\begin{tikzpicture}[scale=0.15]%
    \draw (0,0) circle (1);
    \fill[fill opacity=0.5,fill=blue] (0,0) -- (90:1) arc (90:90-#1*3.6:1) -- cycle;
    \end{tikzpicture}}

\usepackage{hyperref}
\begin{document}
\title{LIRA-V: Lightweight Remote Attestation for Constrained RISC-V Devices} %\vspace{-0.5cm}

%\author{Authors Anonymised for Review}
% affiliation
\author{\IEEEauthorblockN{Carlton Shepherd}
\IEEEauthorblockA{Smart Card and IoT Security Centre\\
Information Security Group\\
Royal Holloway, University of London\\
Egham, Surrey, United Kingdom\\
carlton.shepherd@rhul.ac.uk}
\and
\IEEEauthorblockN{Konstantinos Markantonakis}
\IEEEauthorblockA{Smart Card and IoT Security Centre\\
Information Security Group\\
Royal Holloway, University of London\\
Egham, Surrey, United Kingdom\\
k.markantonakis@rhul.ac.uk}
\and
\IEEEauthorblockN{Georges-Axel Jaloyan}
\IEEEauthorblockA{DIENS\\\'{E}cole Normale Sup\'{e}rieure\\CNRS, PSL University\\
Paris, France}}


%Telephone: (800) 555--1212, Fax: (888) 555--1212}


\maketitle

\begin{abstract}
This paper presents {{LIRA-V}}, a lightweight system for performing remote attestation between constrained devices using the RISC-V architecture. We propose using read-only memory and the RISC-V Physical Memory Protection (PMP) primitive to build a trust anchor for remote attestation and secure channel creation. Moreover, we show how LIRA-V can be used for trusted communication between two devices using mutual attestation.  We present the design, implementation and evaluation of LIRA-V using an off-the-shelf {RISC-V} microcontroller and present performance results to demonstrate its suitability. To our knowledge, we present the first remote attestation mechanism suitable for constrained RISC-V devices, with applications to cyber-physical systems and Internet of Things (IoT) devices.
\end{abstract}

% \fi
\IEEEpeerreviewmaketitle

\section{Introduction}
Constrained devices, e.g., microcontroller units (MCUs), are used ubiquitously in home and building automation, manufacturing, assistive technologies, and industrial control systems. In these situations, it is important to prevent damaging behaviour arising from malware and other unauthorised software. To this end, remote attestation (RA) protocols have been developed to assess remote devices for the presence of unauthorised software. RA is a challenge-response protocol where a remote verifier, $\mathcal{V}$, ascertains the platform operating state of a proving device, $\mathcal{P}$. RA systems rely on a trusted component---a trust anchor---for measuring and signing response reports, or \emph{quotes}, from $\mathcal{V}$.  Increasingly, constrained devices are equipped with CPU security extensions from which trust anchors can be built with hardware-assisted secure execution. %RA is a challenge-response protocol that enables a remote verifier, $\mathcal{V}$, to ascertain the platform configuration state of a target proving device, $\mathcal{P}$.% For example, by $\mathcal{P}$ securely collecting software and firmware measurements in order for $\mathcal{V}$ to detect the presence of unauthorised, potentially malicious components that deviate from known measurements.

Introduced in 2010, RISC-V is a major architectural divergence from popular proprietary architectures, such as Intel and ARM. Consequently, their associated security technologies, e.g., Intel SGX and ARM TrustZone respectively, cannot be readily used for creating RA trust anchors on RISC-V devices. Meanwhile, other traditional roots of trust, namely the Trusted Platform Module (TPM) and hardware secure elements, are generally considered too cumbersome for MCU-type devices~\cite{eldefrawy2012smart,brasser2015tytan,eldefrawy2017hydra,shepherd2017emlog}. 

RISC-V RA mechanisms have been developed for high-end RISC-V devices using CPU protection modes to implement a privileged security monitor (as used by Keystone~\cite{lee2020keystone} and Lebedev et al.~\cite{lebedev2018secure}). However, research has not addressed constrained RISC-V devices that have limited computing power, may not possess a memory management unit (MMU), and may only use a single CPU protection mode.
%the standardisation process of RISC-V TEEs remains ongoing via the RISC-V Trusted Execution Environment Task Group.

This paper presents the first RA mechanism for constrained RISC-V devices. We leverage the Physical Memory Protection (PMP) primitive~\cite{riscv:privileged_arch} and read-only memory (ROM) to build a trust anchor that does not require a fully-fledged TEE, separate CPU privilege modes, or dedicated security hardware, like TPMs and secure elements. Specifically, we leverage PMP to create execute-only memory as a lightweight method for protecting quote signing key secrecy, while device ROM ensures the integrity of RA measurement and reporting procedures. With this, we design and develop a new remote attestation mechanism, and describe its use in secure and trusted device-to-device communication. To support this, we develop a bi-directional mutual attestation protocol, which is verified using formal symbolic analysis. LIRA-V is implemented and evaluated using an off-the-shelf RISC-V MCU, which executes in $\sim$11--32s for secure and trusted channel creation when attesting 64--256KB of memory on both devices. The average case occupies $<$4KB RAM, thus demonstrating suitability for IETF Class 1 and 2 constrained devices~\cite{bormann2014terminology}.
%---a SiFive Hifive1 Rev B---with a Freedom E31 system-on-chip (300MHz CPU, 4MB flash memory, 16KB SRAM). 
%\begin{itemize}
%    \item We present the first remote attestation mechanism tailored for constrained RISC-V devices, which uses the PMP primitive for securely measuring and communicating measurements between the proving device and a remote verifier.
%    \item Our proposed mechanism provides both uni- and bi-directional trust assurances using traditional and mutual remote attestation respectively. The supporting cryptographic protocols are subjected to formal symbolic verification using the \textsc{Scyther} formal verification tool, which found no attacks.
%    \item  %We freely release a reference implementation as MIT-licensed open-source software for further research by the community\footnote{URL ANONYMISED FOR REVIEW.}.
%\end{itemize}

\section{RISC-V Preliminaries}
%\label{sec:riscv-arch}
\label{sec:pmp}
RISC-V is an open-source, load-store instruction set architecture (ISA) with variable word widths (32, 64 and 128 bits) that enables vendors to pursue modular chip design. Devices must support the \emph{base} ISA with 32-bit integer operations as a minimum requirement. Vendors may implement various extensions thereafter, such as single and double-precision floating-point, hypervisor, vector, and single-instruction multiple data instructions. Cryptographic and trusted execution environment (TEE) extensions are under development at the time of writing.

The RISC-V Privileged Architecture Specification~\cite{riscv:privileged_arch} describes three CPU privilege modes: \emph{machine} (M), \emph{supervisor} (S) and \emph{user} (U) mode. M mode is the highest privileged level, akin to EL3 in ARMv8, that is intended for device firmware, e.g., bootloaders. Machine mode is uninterruptible, cannot be disrupted by lower privilege levels, and is the first mode entered upon a device reset. M mode also operates only on physical addresses; it does not access memory using virtual address translation. All RISC-V implementations must support M mode as a mandatory requirement. S-mode is designed for privileged OS services, such as virtual memory management, while U-mode is intended for low privilege applications. In general, embedded devices---the focus of this work---are envisaged to possess either only M or M and U modes~\cite{riscv:privileged_arch}.

\begin{figure}
    \centering
    \includegraphics[width=0.94\linewidth,interpolate=on]{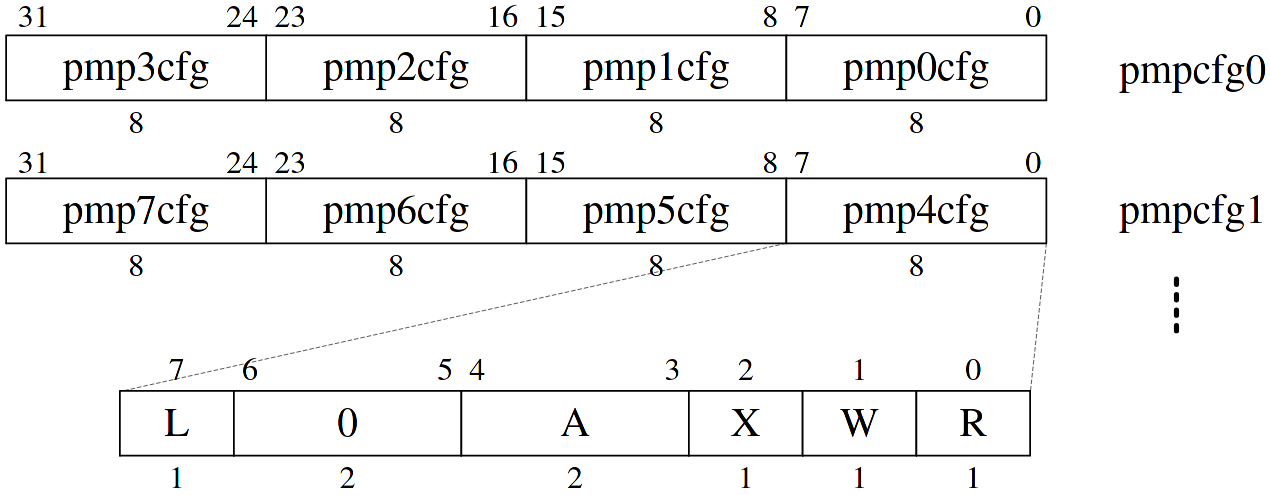}
    \caption{RISC-V PMP configuration registers~\cite{riscv:privileged_arch}.}
    \label{fig:pmp}
    \vspace{-0.1cm}
\end{figure}

Physical Memory Protection (PMP) is a RISC-V security enhancement proposed in 2017 that allows the designation of access control permissions---read (\texttt{R}), write (\texttt{W}), execute (\texttt{X})---to physical memory address regions while in machine mode. PMP is configured using a bank of RISC-V control status registers (CSRs) that specify the desired permission and the associated address (\texttt{A}) region. Setting the PMP lock (\texttt{L}) bit can be optionally used to prevent M-mode from modifying that region's access control permissions once set; locked registers are released only when the CPU is reset. PMP checks are applied to all memory accesses when the current processor privilege mode is in S- or U-mode when the \texttt{L} is not set, or to all modes when \texttt{L} is set. PMP access violations are trapped at the processor level. Setting target regions and their access control parameters is performed using the PMP address (\texttt{pmpaddr}) and configuration (\texttt{pmpcfg}) registers on a supported RISC-V device, shown in Fig. \ref{fig:pmp}.

\section{Remote Attestation and Related Work}

Traditional RA involves a remote verifier, $\mathcal{V}$, that requests the operating state of a single target device, $\mathcal{P}$. At a high level, RA systems comprise \emph{measurement} and \emph{reporting} stages. Historically, the TPM developed by the Trusted Computing Group (TCG) has served as a popular root of trust for storing system measurements. The TPM computes a cryptographically secure hash function over critical system component binaries and stores these measurements in its platform configuration registers (PCRs). During the reporting phase, the PCR values are signed using a hardware-bound attestation identity key (AIK) to produce the quote report that is sent $\mathcal{P}\to\mathcal{V}$. 

RA protocols have been proposed using various software and hardware trust anchors, such as physically unclonable functions (PUFs)~\cite{lebedev2018secure,akram2012privacy}; TEEs, e.g., Intel SGX and ARM TrustZone~\cite{schulz2017boot,anati2013innovative,shepherd2017establishing}; and custom hardware for read-only and guaranteed execution (TyTan~\cite{brasser2015tytan}, SMART~\cite{eldefrawy2012smart}, VRASED~\cite{nunes2019vrased}, TrustLite~\cite{koeberl2014trustlite}, HYDRA~\cite{eldefrawy2017hydra}). Comparing known measurements at regular time intervals (ERASMUS~\cite{carpent2018erasmus}), traversing control-flow integrity (CFI) graphs~\cite{nunes2020tiny,abera2016c,dessouky2017fat,dessouky2018litehax,zeitouni2017atrium,sun2020oat}, and using execution time as an out-of-band channel (SWATT~\cite{seshadri2004swatt} and Pioneer~\cite{seshadri2005pioneer}) have also been explored. %Recent work has also proposed measuring the nodes of a target program's control-flow integrity (CFI) graph, rather than integrity measurement of memory regions or software binaries.

%After receiving an RA request, $\mathcal{P}$ measures security-critical device software components, which may be conducted at boot-time or run-time, and signs the response using the private component of a key-pair, which is verified by $\mathcal{V}$ using the corresponding public key.

\begin{figure}
    \centering
    \includegraphics[width=\linewidth]{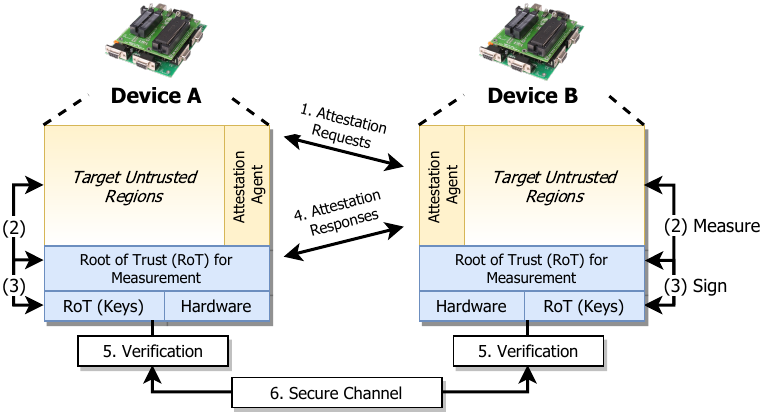}
    \caption{Bi-directional or mutual remote attestation.}
    \label{fig:mutual_ra}
    \vspace{-0.3cm}
\end{figure}

These schemes address attestation involving a single $\mathcal{P}$ and $\mathcal{V}$. Here, $\mathcal{V}$ transmits an attestation request to $\mathcal{P}$, which measures the state of a single or multiple software components at boot- or run-time. The measurement quote is signed using a private key accessible only to the trust anchor, which is returned to and verified by $\mathcal{V}$ using the corresponding public key. In contrast, an emerging paradigm is the development of proposals that attest multiple devices in a single protocol instance. \emph{Mutual} or \emph{bi-directional} RA has been proposed for attesting and bootstrapping secure channels between two devices---shown in Fig.~\ref{fig:mutual_ra}---using ARM TrustZone \cite{shepherd2017establishing,shepherd2018remote}, TPMs~\cite{greveler2011mutual,gasmi2007beyond}, and PUFs as trust anchors~\cite{akram2012privacy}. In other work, the Seda~\cite{asokan2015seda} and SANA~\cite{ambrosin2016sana} systems address RA of device swarms by constructing an efficient topological path between its constituents and aggregating the responses for $\mathcal{V}$. The reader is referred to Ambrosin et al.~\cite{ambrosin2020collective} for a recent comprehensive survey of collective attestation schemes.

\section{Proposed Measurement Procedure}
\label{sec:design}
This work contributes to the area of mutual remote attestation by proposing a novel mechanism using ROM and RISC-V PMP as a trust anchor. The following sections describe the threat model and system design in detail.

\subsection{Threat Model and Design Goals}
\label{sec:design_goals}

We consider a privileged software adversary that has read, write and execute access to memory that is not explicitly protected by hardware-assisted access control, i.e. PMP and read-only memory. Such an adversary may exploit an existing software vulnerability---a vulnerable network service or connected I/O device---that provides access unprotected memory regions. The attacker then establishes a persistent presence on the device by overwriting untrusted memory.   The attacker may operate in M mode, which may be the only CPU protection mode supported by a constrained device. However, in line with other hardware-based RA proposals, we \emph{do} trust the RISC-V PMP primitive and a ROM unit for preserving the integrity of signing keys and the core root of trust for measurement (CRTM) code. As such, sophisticated attacks that bypass the protections afforded by PMP, like physical fault injections, are considered beyond the scope of this work. Additionally, we assume that the measuring procedure executes atomically. Numerous proposals have been made for enforcing this \cite{francillon2014minimalist,carpent2018reconciling} using a small hardware extension for validating program counter (PC) bounds to enforce code entry points. Unlike existing proprietary processor architectures, this is significantly eased by the open-source nature of many RISC-V cores. 

The following security goals are addressed in this work:
\begin{enumerate}[start=1,label={[G\arabic*]}]
    \item \textbf{Measurement Procedure Integrity}: the proposal shall resist privileged attacks that attempt to modify the code for measuring the device's current operating state.
    \item \textbf{Signing Key Secrecy}: the key used for signing response quotes for entity authentication shall remain secret against privileged adversaries under the threat model.
    \item \textbf{Secure Channel Creation}: the proposed mechanism shall bootstrap a secure channel for bi-directional data transfer between $\mathcal{P}$ and $\mathcal{V}$ with mutual entity authentication, replay protection, and forward secrecy. 
    \item \textbf{Bi-Directional Attestation}: secure channel creation shall be predicated on $\mathcal{P}$ and $\mathcal{V}$ having known and authorised platform configuration states in a single protocol instance.  
\end{enumerate}

The following deployability goals are also considered:
\begin{enumerate}[start=1,label={[D\arabic*]}]
    \item \textbf{Constrained Device Suitability}: the proposed system shall be suitable for constrained devices, e.g., MCUs. We assume the devices can execute standard public-key cryptographic algorithms, but can benefit from lightweight mechanisms due to computational constraints. This corresponds to IETF Class 1 and 2 devices (10--50KB RAM, 100--250KB storage, and a CPU in the MHz range)~\cite{bormann2014terminology}.
    \item \textbf{Cost and Flexibility}: the proposal shall avoid relying upon external security hardware or CPU extensions, such as privilege modes unlikely to exist on embedded devices, e.g., hypervisor modes, and security co-processors.
\end{enumerate}

\subsection{Proposed Measurement Procedure}
\label{sec:measurement_procedure}

The proposed trust anchor is underpinned by two trusted components: the RISC-V PMP primitive for configuring physical memory regions with access control permissions (see \S\ref{sec:pmp}) and ROM code for configuring PMP registers at start-up. The target is a standard, single-core embedded system with its components deployed in a system-on-chip (SoC) that executes basic configuration routines in ROM before transferring control to the primary program in persistent memory. LIRA-V comprises the following stages, shown in Fig.~\ref{fig:ra_execution_Flow}:

\begin{figure}
    \centering
    \includegraphics[width=0.9\linewidth]{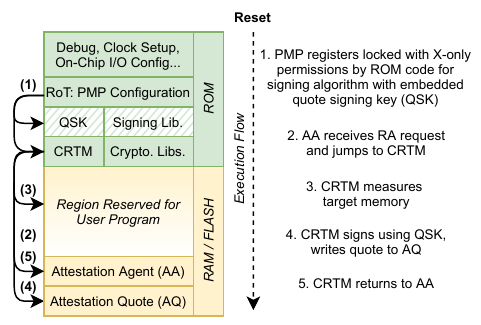}
    \caption{Remote attestation measurement procedure. Trusted components shown in green, untrusted in yellow, and PMP-configured regions are hatched.}
    \label{fig:ra_execution_Flow}
    \vspace{-0.3cm}
\end{figure}

\subsubsection{PMP Configuration}
Following a reset signal and basic SoC setup, ROM code is used to assign PMP access control permissions to physical memory regions. A single PMP entry is configured to denote execute-only (X-only) permissions to the region that encompasses the quote signing key (QSK) embedded in its associated signing method. For measurement integrity, ROM is used to host the CRTM and its dependent cryptographic algorithms.
\subsubsection{Attestation Request}
The attestation agent (AA) is untrusted code that implements the services for transmitting/receiving attestation messages to/from $\mathcal{V}$ over a network medium, e.g., Wi-Fi, Bluetooth, or LPWAN. AA invokes the ROM CRTM for executing the measurement procedure after receiving an attestation request.
\subsubsection{CRTM}
The ROM CRTM computes an aggregated hash of the device physical memory region of the form $H(\{m_{i}, m_{i+1}, \dots, m_{i+b}\}, H(\{m_{i+1+b}, m_{i+2+b},\dots,m_{i+j+b}\},$\\$H(\dots)))$, where $H(\cdot)$ is a cryptographic hash function, $m_{i}$ is the contents of the $i^{th}$ address, and $b$ is the address block size to be hashed. The attested region is pre-programmed in ROM CRTM that encompasses all or part of the memory map. We evaluate the performance of varying regions in \S\ref{sec:eval}. 
    
\subsubsection{Quote Signing} The aggregated measurement is signed using the ROM QSK for entity authentication. The corresponding verification key is held by $\mathcal{V}$ before deployment. As per [G2] (\S\ref{sec:design_goals}), the QSK must remain confidential to prevent an adversary forging valid quotes while unauthorised software persists on the device. We leverage RISC-V PMP to set X-only memory to a signing procedure with an embedded QSK as a lightweight method for achieving this. Lastly, the signed attestation quote is placed into AA-readable memory. 

\subsubsection{Attestation Response} AA returns the attestation quote (AQ) to $\mathcal{V}$ as part of a secure remote attestation protocol (\S\ref{sec:protocol}), who verifies the signature and whether the quote measurement conforms to expectations.

\section{Proposed Protocol}
\label{sec:protocol}

This section develops the \emph{reporting} phase of LIRA-V for secure quote transmission, which formalises a novel secure channel protocol with bi-directional attestation.

\subsection{Assumptions and Threat Model}

From [G3] and [G4] (\S\ref{sec:design_goals}), the aim is to bootstrap a secure channel following mutual attestation---see Fig. \ref{fig:mutual_ra}---where two constrained devices serve as $\mathcal{P}$ and $\mathcal{V}$. Each device is assumed to implement the measurement procedure outlined in \S\ref{sec:design}. During the protocol, we address network-based adversaries that attempt to forge attestation responses and intercept, replay, redirect and otherwise manipulate protocol messages permitted under the Dolev-Yao adversarial model. We assume that the proving devices can access a cryptographically secure entropy source for random number generation and can execute standard public key cryptography algorithms ([D1], \S\ref{sec:design_goals}).

\subsection{Offline Phase}

Each device is provisioned a unique QSK used during the measurement phase for entity authentication. The corresponding public key and expected measurements for verifying responses are provisioned into the opposing devices in PMP-protected ROM by an administrator. For performance, a compact signature scheme is suggested, e.g., Ed25519 (256-bit key sizes). Devices may also be enrolled in a group signature scheme with multiple QSKs mapped to a single group verification key for scalability purposes. Elliptic Curve Diffie-Hellmann (ECDH) key exchange is proposed as an efficient basis for mutual key agreement. As such, common domain parameters, $\Delta = (p,a,b,G,\eta,h)$ (in the prime case), must be standardised across devices.  The domain parameters, signature verification keys, and expected measurement hashes are provisioned in trusted ROM prior to deployment.

\begin{algorithm}[t!]
	\floatname{algorithm}{Protocol}
	\caption{Proposed Mutual Attestation Protocol}
    
	\begin{algorithmic}[1]
		\STATE $A\rightarrow{B}$ : $n_{A}$ $||$ $q_{A}$ $||$ $AR$ \\
        \vspace{0.16cm}
		\STATE $B\rightarrow{A}$ : $n_{B}$ $||$ $q_{B}$ $||$ $AR$ $||$  $AE_{K}(Q_{B}$ $||$ $\sigma_{B}(H(Q_{B}$ $||$ $n_{A}$ $||$ $n_{B}$ $||$ $q_{B}$ $||$ $q_{A})))$\vspace{0.16cm}

		\STATE $A\rightarrow{B}$ : $AE_{K}(Q_{A}$ $||$ $\sigma_{A}(H(Q_{A}$ $||$ $n_{A}$ $||$ $n_{B}$ $||$ $q_{A}$ $||$ $q_{B}))$
	\end{algorithmic}
	\label{protocol:1}
\end{algorithm}
\begin{table}
\caption{Notation for Protocol~\ref{protocol:1}.}
\resizebox{\linewidth}{!}{%
\def\arraystretch{1}%
\begin{tabular}{ll} 
\toprule
\textbf{Notation} & \textbf{Description} \\\midrule
$D_{ID}$ & Identifier of device $D$, e.g, IP address or UUID.\\\midrule

$n_{D}$ & Cryptographically secure nonce generated by $D$.\\\midrule

$q_{D}$ & \makecell[l]{Public ECDH value generated by $D$ (scalar multiplication\\ of an agreed base and $D$'s randomly generated secret point.)}\\\midrule
$AR$ & Attestation request operation. \\\midrule
$H(\cdot)$ & Cryptographically secure one-way hash function.\\\midrule
$X$ $||$ $Y$ & Concatenation of $X$ and $Y$.\\\midrule
$AE_{K}(\cdot)$ & Authenticated symmetric encryption algorithm keyed with $K$.\\\midrule
$Q_{D}$ & Measurement response quote signed by $D$.\\\midrule
$\sigma_{D}(\cdot)$ & Message signed by $D$ using crytographic signature algorithm.\\\bottomrule
\end{tabular}
}
\label{tab:notation}
\vspace{-0.2cm}
\end{table}

\subsection{Online Phase}
\label{sec:online_phase}
The proposed mutual RA protocol is presented in Protocol \ref{protocol:1}, comprising three messages between devices $A$ and $B$:
\subsubsection{M1}
 $A$ initiates the protocol with $B$ by transmitting a cryptographically secure nonce, $n_A$; its public ECDH point, $q_{A} = d_{A} \cdot G$, where $d_{A}$ is the randomly generated private key and $G$ is the base using domain parameters, $\Delta$; and an attestation request flag, $AR$, of $A \to B$.
 \subsubsection{M2}
 $B$ executes the measurement process in \S\ref{sec:design} to produce its signed attestation quote, $Q_{B}$. $B$ generates its secure nonce, $n_{B}$, and ECDH public point, $q_B = d_{B} \cdot G$. Using $q_A$ from $A$, $B$ derives the shared ECDH secret using $d_{B} \cdot q_{A} \cdot G$, and applies a key derivation function to derive an ephemeral session key, $K$. This key is used with an authenticated encryption algorithm, $AE$, e.g., AES-GCM, to encrypt its quote response, $Q_{B}$, and a signed message of $H(Q_{B} || n_{A}  || n_{B} || q_{B}  || q_{A})$ signed using $B$'s QSK.
\subsubsection{M3}
$A$ computes the shared ECDH secret ($d_{A} \cdot q_{B} \cdot G$) and derives $K$. $A$ decrypts M2 the sent by $B$ and verifies the signature and measurements of $Q_{B}$ therein. $A$ aborts the protocol if $Q_{B}$ fails $A$'s signature verification, the measurements deviate from the expected value, or $AE$ contains an invalid authentication tag.  Next, $A$ executes its measurement procedure, after which $AE_{K}$ is used to encrypt the measured quote, $Q_A$, and $H(Q_{A} || n_{A} || n_{B} || q_{A} || q_{B})$ under $A$'s QSK, which are sent to $B$. $B$ then validates the signature and measurement contents of $A$'s quote using its public key. $\square$

\subsection{Formal Verification}
We subject the proposed bi-directional attestation protocol to formal verification using the \textsc{Scyther} analyser~\cite{cremers2008scyther}, which operates under the symbolic model using the perfect cryptography assumption. Previously, \textsc{Scyther} has been used to formally verify Internet Key Exchange (IKEv1 and IKEv2)~\cite{cremers2011key}, ISO/IEC 9798~\cite{basin2013provably} and 11770~\cite{cremers2014improving} protocol families, and WiMAX (IEEE 802.16)~\cite{taha2009formal}. A given protocol is specified using the Security Protocol Description Language (SPDL) defining the communicating entities (\emph{roles}); the protocol messages using built-in primitives, e.g., nonces, hash functions, symmetric and public-key signature algorithms, and user-defined types; and the security properties to assess (\emph{claims}). \textsc{Scyther} analyses whether the claims hold against all possible behaviours of a Dolev-Yao adversary (\emph{traces}) using the protocol specification.  No attacks were discovered on the proposed protocol; the claims of quote and session key secrecy (confidentiality), reachability, aliveness (weak authentication) and non-injective agreement and synchronisation (replay attack protection) were successfully maintained. The full protocol specification is released freely.\footnote{Protocol \textsc{Scyther} source URL: \url{https://cs.gl/extra/ecc-btp-scyther.spdl}}

\section{Protocol Implementation}

\subsection{Platform Specifications}
We implemented a proof-of-concept of our proposal using the SiFive HiFive1 Rev B---a 32-bit RISC-V single-board computer with a low-powered FE310-G002 MCU SoC with 16KB SRAM and 4MB off-chip SPI flash memory. The SoC hosts a single-core E31 CPU at 320MHz; a 16KB L1 instruction cache; multiplication, atomic, and compressed instruction extensions (RV32IMAC); and support for the Privileged Architecture specification with 8 configurable PMP registers~\cite{riscv:privileged_arch}. 

\subsection{Measurement Phase}

%\begin{listing}
%\begin{minted}[fontsize=\footnotesize]{C}
%#define SHA3_256_LEN 32
%#define BLOCK_SIZE 1024
%static uint8_t agg_hash[SHA3_256_LEN];
%void crtm_phy_mem_measure(
%    uint32_t start_addr,
%    uint32_t end_addr) {
%    uint8_t tmp[SHA3_256_LEN+BLOCK_SIZE];
%    for (uint32_t i=start_addr;
%         i<=end_addr-BLOCK_SIZE;
%         i+=BLOCK_SIZE) {
%      memcpy(tmp, (const uintptr_t *) i, BLOCK_SIZE);
%      memcpy(tmp+BLOCK_SIZE, agg_hash, SHA3_256_LEN);
%      sha3(tmp, SHA3_256_LEN+BLOCK_SIZE,
%           agg_hash, SHA3_256_LEN);
%   }
%}
%\end{minted}
%\caption{Implemented CRTM measurement C code.}
%\label{lst:crtm}
%\end{listing}

The CRTM (\S\ref{sec:measurement_procedure}) enumerates a memory range in contiguous blocks between two pre-programmed addresses. On request, the CRTM function computes an aggregated hash of the memory contents in this range to create the raw measurement. SHA-3 was used as the hash function using \texttt{tiny\_sha3}, a portable implementation of FIPS-202/SHA-3~\cite{tinysha3}.
The Freedom Metal library of the Freedom E SDK---a hardware abstraction layer for providing portability between Freedom E RISC-V targets---was used for PMP configuration. The SDK provides C interfaces for configuring \texttt{pmpcfg} and \texttt{pmpaddr} registers to assign access control permissions, e.g., \texttt{R}/\texttt{W}/\texttt{X}/\texttt{L} (Fig. \ref{fig:pmp}), to a contiguous memory address region.

\subsection{Quote Reporting}
\label{sec:reporting_setup}
The protocol was implemented using the RISC-V port of the Network and Cryptography Library (NaCl) for ECC mutual key agreement using X25519 and symmetric authenticated encryption using the XSalsa20-Poly1305 construction~\cite{riscv:nacl,bernstein2011nacl}. SHA-3 served as the CRTM hash function using \texttt{tiny\_sha3}, an embedded implementation of FIPS-202/SHA-3~\cite{tinysha3}. For QSK, Ed25519 was employed using an adapted implementation from the Keystone project~\cite{lee2020keystone}. This adaption replaced using a byte buffer for defining the signing key with an assembly routine that loaded immediate values from PMP-protected static memory for X-only execution. Both devices were connected to a Windows workstation over USB serial for inter-device connectivity and protocol orchestration. A Python script was developed for benchmarking runs using the Time module with microsecond precision, and used the PySerial module for initiating the protocol on device $A$ over the serial interface. The timer was terminated by the script after reading an acknowledgement serial message from device $B$ after it received and validated [M3] (\S\ref{sec:online_phase}).

\section{Evaluation}
\label{sec:eval}
\subsection{Performance Analysis}

\begin{figure}
    \centering
    \includegraphics[width=0.9\linewidth]{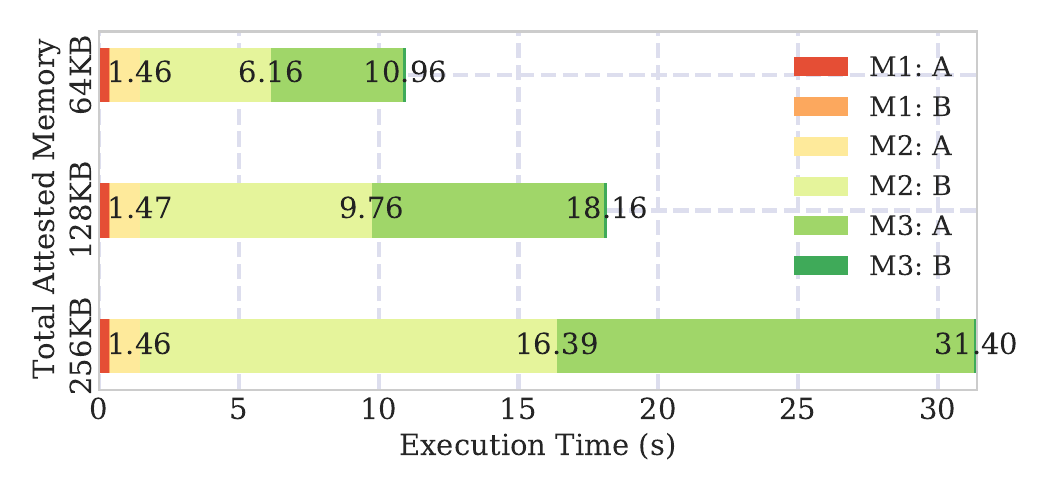}\vspace{-0.4cm}
    \caption{Mean cumulative protocol execution time.}
    \label{fig:total_time}
    \vspace{-0.2cm}
\end{figure}

Our implementation was cross-compiled for 32-bit RISC-V platforms using the GNU C Compiler with the \texttt{-0s} optimisation flag, and deployed on both boards using the Freedom E SDK~\cite{freedomesdk}. The stack and heap were globally limited to use 6KB and 1KB respectively per device. We first benchmarked the total protocol execution time and its distribution by message and entity. Here, the protocol was executed 100 times using the setup in \S\ref{sec:reporting_setup} for total attested memory ranges between 64KB--256KB. The mean execution times were computed for each message and by device. The results are presented in Table \ref{tab:protocol_msg_times} and Fig. \ref{fig:total_time}. The measurements show that [M2] and [M3] dominate the total execution time, accounting for $\sim$48\% per message of the total protocol time. 

\begin{table}
\centering
\caption{Mean protocol wall-clock times (seconds) by device for various attested memory ranges (1KB blocks, 100 itrs.; 4.S.F).}
\label{tab:msgs}
\resizebox{\linewidth}{!}{%
\begin{tabular}{c|ccc|ccc|ccc}
\toprule
& \multicolumn{9}{c}{\textbf{Total CRTM Attested Memory}}\\
 & \multicolumn{3}{c}{\textit{64KB}} &  \multicolumn{3}{|c}{\textit{128KB}} &  \multicolumn{3}{|c}{\textit{256KB}} \\\midrule
\textbf{Message} &  $A$ & $B$ & \textbf{Total} &  $A$ & $B$ & \textbf{Total}  &  $A$ & $B$ & \textbf{Total} \\\midrule
\textbf{M1:} $A\to B$ & 0.342 & 0.049 & 0.391  & 0.346 & 0.052 & 0.398 & 0.341 & 0.048 & 0.389\\
\textbf{M2:} $B \to A$ & 1.073 & 4.698 & 5.771 &1.069 & 8.289  & 9.358 & 1.067 &  14.93& 16.00\\
\textbf{M3:} $A \to B$ & 4.701 & 0.097 & 4.798 & 8.302 & 0.103 & 8.405 & 14.92 & 0.090 & 15.01 \\\midrule
\textbf{Grand Total} & 6.116 & 4.844 & 10.96 & 9.717 & 8.444 &  18.16 & 16.33 & 15.07 & 31.40 \\\bottomrule
\end{tabular}
}
\label{tab:protocol_msg_times}
\end{table}

\begin{table*}
\centering
\caption{Mean wall-clock CRTM execution time (seconds) for varying memory block sizes and total memory (100 itrs.; 4 S.F.).}
\label{tab:results1}
%\resizebox{\linewidth}{!}{%
\begin{tabular}{@{}r|ccccccccccccc@{}}
\toprule
      \textbf{Block}         &      \textbf{1KB} & \textbf{2KB} & \textbf{4KB} & \textbf{8KB} &\textbf{16KB} & \textbf{32KB} & \textbf{64KB} & \textbf{128KB} & \textbf{256KB} & \textbf{512KB} & \textbf{1MB} & \textbf{2MB} & \textbf{4MB} \\ \midrule
\textbf{1KB} &    0.050 & 0.102  &  0.206   &  0.411   &  0.826 &   1.646        &   3.311       &  6.625  &    13.25       & 26.53    & 54.76   &  103.5       &   207.2    \\
\textbf{2KB}   & ---  & 0.099 & 0.202       &    0.409       &  0.823  &  1.650  &   3.309          &   6.621          &  13.24    &  26.41  & 53.71 &   102.6           &    204.0      \\
\textbf{4KB}   &  --- & --- &   0.191  &   0.398         &  0.802  &    1.609   &    3.212       &      6.431          &   12.86  & 25.53 &  50.27 &  100.5      &  201.1     \\\bottomrule
\end{tabular}%
%}
\label{tab:crtm_measurement_time}
\vspace{-0.2cm}
\end{table*}

\begin{figure}
    \centering
    \includegraphics[width=0.9\linewidth]{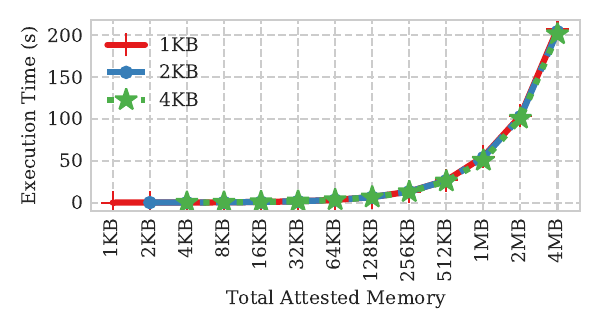}\vspace{-0.4cm}
    \caption{CRTM execution time for varying attested memory and block sizes.}
    \label{fig:crtm_time}
    \vspace{-0.2cm}
\end{figure}

From this, we further characterised CRTM measurement time against the attested memory size and the block size (the number of contiguous memory addresses that are hashed). The default block size was tested at values between 1KB---4KB, the maximum our test devices could accommodate,\footnote{The global stack required 6KB when using 4KB blocks, while only a 4KB stack was needed to support 1KB blocks in our implementation.} while the total attested memory ranged from 1KB to the maximum size of flash memory (4MB). These results are shown in Table \ref{tab:crtm_measurement_time} and Fig. \ref{fig:crtm_time}.  Expectedly, the measurement time is directly proportional to the total attested memory. At worst, attesting 4MB required 207.2s (1KB blocks) and 201.1s (4KB). In general, the rate of increase is $\sim$1.6s for every 32KB of attested memory, while the total time doubles for each doubling of attesting memory. A 1KB--4KB increase in block size produces a small reduction ($\sim$3\%) of the measurement time, thus exhibiting a memory-time trade-off given the greater required stack allocation for 4KB blocks.

\subsection{Security Evaluation, Attacks, and Countermeasures}

The threat model in \S\ref{sec:design_goals} describes a privileged adversary that can access memory that is not protected by hardware security. In \S\ref{sec:design}, we proposed moving the root of trust to ROM, which executes immediately following a device reset and before the execution of untrusted memory. This configures PMP regions to secure the integrity of the CRTM, thus satisfying [G1]. The confidentiality of QSKs is preserved by embedding the key into an X-only signing routine configured in start-up ROM using PMP with the \texttt{L}-bit set, thus meeting [G2].  We also proposed a novel protocol that bootstraps a secure channel using ephemeral keys and the mutual attestation of both devices for meeting [G3] and [G4].
We now analyse some specific attacks under the threat model in \S\ref{sec:design_goals}:

\begin{enumerate}[start=1,label={[A\arabic*]}]
    \item \emph{Modifying QSK's PMP address range or its access control permissions}. The QSK PMP entry is locked after being set by start-up ROM, which prevents writes to its configuration (\texttt{pmpcfg}) and address (\texttt{pmpaddr}) registers in M mode. Locked entries are released only after a device reset; since PMP registers are configured \emph{before} untrusted memory is afforded control, the attacker is unable to modify QSK's PMP registers after boot-time.
  \item \emph{Transferring removable flash memory.} Many embedded devices rely on removable memory, e.g., SD cards. The trust anchor---QSK and PMP configuration code---resides in ROM, and so replacing removable memory would affect attestation agent and signed quote availability. The device would be unable to respond to RA requests, akin to disabling the device. Such attacks are difficult to address without non-removable memory or tamper-resistant casing. 
  \item \emph{Adversary overwrites the attestation quote (AQ) in untrusted memory.} This yields effects akin to [A2]: the device will fail to transmit a valid quote to $\mathcal{V}$, therefore failing the attestation process and notifying $\mathcal{V}$ of an issue.
  \item \emph{Leveraging side-effects of X-only code.}  We use RISC-V PMP to assign an X-only region to protect the confidentiality of quote signing. A general limitation of X-only code is that, while direct reads and writes are prohibited, CPU caches and registers still exhibit the effects of executed code. Consequently, these must be correctly sanitised to mitigate potential information leakage before returning control flow to untrusted memory. 
\end{enumerate}

\subsection{Limitations}

Our proposal focused on single-core MCUs that possess a single application and CPU protection mode. Attesting dynamically loaded tasks or applications using a real-time OS (RTOS) poses greater challenges, which we defer to future research. Secondly, by relaxing the atomicity requirement, we are able to construct a lightweight and portable system that requires no additional security hardware, e.g., TPMs, or CPU protection modes.  As such, our proposal is not best suited against sophisticated adaptive malware that transiently and reactively relocates during the attestation process to avoid detection. Addressing this challenge remains an open challenge in the literature without resorting to additional hardware~\cite{francillon2014minimalist,eldefrawy2012smart,nunes2019vrased}, which contravenes [D1] (\S\ref{sec:design_goals}). Both Nunes et al.~\cite{nunes2019vrased} and Francillon et al.~\cite{francillon2014minimalist} propose countermeasures using dedicated hardware extensions that enforce controlled invocation and PC validation. In future work, we aim to investigate the use of PMP and trusted ROM to provide these properties. Furthermore, we consider our proof-of-concept performance measurements baseline results: optimisation was not the primary objective, and it is likely that significant benefits could be made through using, for example, a hardware-accelerated hash function for the CRTM.  Lastly, a compromised device may refuse to participate in RA protocols by dropping messages to/from $\mathcal{V}$. Such non-participation attacks are generally beyond the scope of RA systems. A countermeasure is fixing RA timeouts and blacklisting or investigating unresponsive devices.

\section{Conclusion}

\begin{table}
\centering
\renewcommand\arraystretch{0}
\caption{Comparison of related RA proposals using criteria from \S\ref{sec:design_goals}.}
\label{tab:comparison}
\begin{threeparttable}
\resizebox{\linewidth}{!}{%
\begin{tabular}{@{}ccccccccc@{}}
\toprule
\textbf{Root of Trust} & \textbf{Work} & \textbf{G1} & \textbf{G2} & \textbf{G3} & \textbf{G4} & \textbf{D1} & \textbf{D2} \\\midrule
\multirow{3}{*}{TPM} & Gasmi et al.~\cite{gasmi2007beyond} & \priority{100}  & \priority{100} & \priority{100} & \priority{100} & \priority{50} & \priority{0}  \\
& Greveler et al.~\cite{greveler2011mutual} & \priority{100}  & \priority{100} & \priority{0} & \priority{100} & \priority{50} & \priority{0} \\
 & TPM 2.0 DAA~\cite{brickell2004direct} & \priority{100} & \priority{100} & \priority{0} & \priority{0} & \priority{50} & \priority{0} \\\midrule
PUF & Lebedev et al.~\cite{lebedev2018secure} & \priority{100}  & \priority{100} & \priority{0} & \priority{0} & \priority{0} & \priority{50}  \\\midrule
Intel SGX & EPID~\cite{johnson2016intel} & \priority{100} & \priority{100} & \priority{100} &  \priority{0}& \priority{0} & \priority{50} \\\midrule

HW+SW & SMART~\cite{eldefrawy2012smart} & \priority{100} & \priority{100} & \priority{0} & \priority{0} & \priority{100} & \priority{50} \\

Co-Design & VRASED~\cite{nunes2019vrased} & \priority{100} & \priority{100} & \priority{0} & \priority{0} & \priority{100} & \priority{50} \\\midrule

\multirow{2}{*}{EA-MPU} & TyTAN~\cite{brasser2015tytan} & \priority{100} & \priority{100} & \priority{0} & \priority{0} & \priority{100} & \priority{50} \\

& TrustLite~\cite{koeberl2014trustlite} & \priority{100} & \priority{100} & \priority{0} & \priority{0} & \priority{100} & \priority{50} \\\midrule

Microkernel & HYDRA~\cite{eldefrawy2017hydra} & \priority{100} & \priority{100} & \priority{0} & \priority{0} & \priority{50} & \priority{50} \\\midrule

GP TEE & Shepherd et al.~\cite{shepherd2017establishing} & \priority{100} & \priority{100} & \priority{100} & \priority{100} & \priority{50} & \priority{50} \\\midrule
ARM TrustZone & Schulz et al.~\cite{schulz2017boot} & \priority{100} & \priority{100} & \priority{0} & \priority{0} & \priority{100} & \priority{100} \\\midrule
\multirow{2}{*}{RISC-V PMP} & Keystone~\cite{lee2020keystone} &  \priority{100} & \priority{100} &  \priority{0} & \priority{0} & \priority{0} & \priority{100} \\
 & \textbf{LIRA-V} & \priority{100} & \priority{100} &\priority{100}  & \priority{100} & \priority{100} & \priority{100} \\\bottomrule
\end{tabular}%
}
\begin{tablenotes}
\item EA-MPU: Execution-aware memory protection unit, GP: GlobalPlatform.
\item \priority{100}: Satisfies, \priority{50}: Partially satisfies, \priority{0}: Unsatisfactory.
\end{tablenotes}
\end{threeparttable}
\vspace{-0.2cm}
\end{table}
This paper presented LIRA-V, a lightweight attestation system for RISC-V constrained devices. Our proposal uses on-board ROM and PMP to build X-only memory for preserving the integrity and confidentiality of attestation measurement and reporting. To the best of our knowledge, LIRA-V is the first remote attestation mechanism for RISC-V constrained devices that does not require dedicated security hardware, TEEs, or separate CPU privilege modes to build a trust anchor. We also went beyond related work (Table \ref{tab:comparison}) and showed how LIRA-V can support secure device-to-device communication with bi-directional attestation. We presented a multi-part evaluation with performance measurements from an implementation on an off-the-shelf RISC-V MCU, alongside a security analysis and a limitations discussion. Future work includes, once standardised, migrating and evaluating post-quantum cryptography for constrained device attestation.

\section*{Acknowledgments}
 The authors would like to thank Blake Loring and the IEEE SafeThings reviewers for their helpful reviews towards improving this paper. This work has received funding from the European Union's Horizon 2020 research and innovation programme under grant agreement No. 883156 (EXFILES).
%This work is s

%We would like to thank Blake Loring for 
%Anonymised for review.



\bibliographystyle{IEEEtran}
\bibliography{IEEEabrv, bib}


\end{document}